\documentclass[twocolumn]{aastex62}
\graphicspath{{./}{figures/}}

\shorttitle{Off the btfr: a population of baryon-dominated udg\small s}
\shortauthors{Pavel E. Mancera Pi\~na et al.}

\begin{document}

\title{\Large Off the baryonic Tully-Fisher relation: a population of baryon-dominated ultra-diffuse galaxies}

\author[0000-0001-5175-939X]{\large Pavel E. Mancera Pi\~na}
\affiliation{Kapteyn Astronomical Institute, University of Groningen, Landleven 12, 9747 AD, Groningen, The Netherlands}
\affiliation{ASTRON, Netherlands Institute for Radio Astronomy, Postbus 2, 7900 AA Dwingeloo, The Netherlands}

\author[0000-0002-0447-3230]{\large Filippo Fraternali}
\affiliation{Kapteyn Astronomical Institute, University of Groningen, Landleven 12, 9747 AD, Groningen, The Netherlands}

\author[0000-0002-9798-5111]{\large Elizabeth A. K. Adams}
\affiliation{ASTRON, Netherlands Institute for Radio Astronomy, Postbus 2, 7900 AA Dwingeloo, The Netherlands}
\affiliation{Kapteyn Astronomical Institute, University of Groningen, Landleven 12, 9747 AD, Groningen, The Netherlands}

\author[0000-0002-5655-6054]{\large Antonino Marasco}
\affiliation{Kapteyn Astronomical Institute, University of Groningen, Landleven 12, 9747 AD, Groningen, The Netherlands}
\affiliation{ASTRON, Netherlands Institute for Radio Astronomy, Postbus 2, 7900 AA Dwingeloo, The Netherlands}

\author[0000-0002-0616-6971]{\large Tom Oosterloo}
\affiliation{ASTRON, Netherlands Institute for Radio Astronomy, Postbus 2, 7900 AA Dwingeloo, The Netherlands}
\affiliation{Kapteyn Astronomical Institute, University of Groningen, Landleven 12, 9747 AD, Groningen, The Netherlands}

\author[0000-0001-9857-7788]{\large Kyle A. Oman}
\affiliation{Kapteyn Astronomical Institute, University of Groningen, Landleven 12, 9747 AD, Groningen, The Netherlands}

\author[0000-0001-8849-7987]{\large Lukas Leisman}
\affiliation{Department of Physics and Astronomy, Valparaiso University, 1610 Campus Drive East, Valparaiso, IN 46383, USA}

\author[0000-0003-4019-0673]{\large Enrico M. di Teodoro}
\affiliation{Research School of Astronomy and Astrophysics - The Australian National University, Canberra, ACT, 2611, Australia}

\author[0000-0001-9072-5213]{\large Lorenzo Posti}
\affiliation{Universit\'e de Strasbourg, CNRS UMR 7550, Observatoire astronomique de Strasbourg, 11 rue de l'Universit\'e, 67000 Strasbourg, France}

\author[0000-0002-3501-8396]{\large Michael Battipaglia}
\affiliation{Department of Physics and Astronomy, Valparaiso University, 1610 Campus Drive East, Valparaiso, IN 46383, USA}

\author[0000-0002-1821-7019]{\large John M. Cannon}
\affiliation{Department of Physics \& Astronomy, Macalester College, 1600 Grand Avenue, Saint Paul, MN 55105, USA}

\author[0000-0002-2492-7973]{\large Lexi Gault}
\affiliation{Department of Physics and Astronomy, Valparaiso University, 1610 Campus Drive East, Valparaiso, IN 46383, USA}

\author[0000-0001-5334-5166]{\large Martha P. Haynes}
\affiliation{Cornell Center for Astrophysics and Planetary Science, Space Sciences Building, Cornell University, Ithaca,  NY 14853, USA}

\author[0000-0001-9165-8905]{\large Steven Janowiecki}
\affiliation{University of Texas, Hobby-Eberly Telescope, McDonald Observatory, TX 79734, USA}

\author[0000-0000-0000-0000]{\large Elizabeth McAllan}
\affiliation{Department of Physics and Astronomy, Valparaiso University, 1610 Campus Drive East, Valparaiso, IN 46383, USA}

\author[0000-0000-0000-0000]{\large Hannah J. Pagel}
\affiliation{Department of Astronomy, Indiana University, 727 East Third Street, Bloomington, IN 47405, USA}

\author[0000-0001-8530-7543]{\large Kameron Reiter}
\affiliation{Department of Physics and Astronomy, Valparaiso University, 1610 Campus Drive East, Valparaiso, IN 46383, USA}

\author[0000-0001-8283-4591]{\large Katherine L. Rhode}
\affiliation{Department of Astronomy, Indiana University, 727 East Third Street, Bloomington, IN 47405, USA}

\author[0000-0001-8483-603X]{\large John J. Salzer}
\affiliation{Department of Astronomy, Indiana University, 727 East Third Street, Bloomington, IN 47405, USA}

\author[0000-0002-3222-2949]{\large Nicholas J. Smith}
\affiliation{Department of Astronomy, Indiana University, 727 East Third Street, Bloomington, IN 47405, USA}

\correspondingauthor{Pavel E. Mancera Pi\~na}
\email{pavel@astro.rug.nl}

\nocollaboration



\begin{abstract}
We study the gas kinematics traced by the 21-cm emission of a sample of six H\,{\sc i}--rich low surface brightness galaxies classified as ultra-diffuse galaxies (UDGs). Using the 3D kinematic modelling code $\mathrm{^{3D}}$Barolo we derive robust circular velocities, revealing a startling feature: H\,{\sc i}--rich UDGs are clear outliers from the baryonic Tully-Fisher relation, with circular velocities much lower than galaxies with similar baryonic mass. Notably, the baryon fraction of our UDG sample is consistent with the cosmological value: these UDGs are compatible with having no ``missing baryons" within their virial radii. Moreover, the gravitational potential provided by the baryons is sufficient to account for the amplitude of the rotation curve out to the outermost measured point, contrary to other galaxies with similar circular velocities. We speculate that any formation scenario for these objects will require very inefficient feedback and a broad diversity in their inner dark matter content.
\end{abstract}



\section{Introduction} 

The baryonic Tully-Fisher relation (BTFR; \citealt{mcgaugh1, mcgaugh2}) is a tight sequence in the baryonic mass--circular velocity plane followed by galaxies of different types (e.g. \citealt{heijer}; \citealt{lelli2016}; \citealt{anastasia}). It has been of paramount importance and widely used for calibrating distances to extragalactic objects and to constrain, for example, semi-analytical and numerical models of galaxy formation and evolution (e.g. \citealt{governato}; \citealt{dutton}; \citealt{mcgaugh3}; \citealt{sales}, and references therein).


Among the galaxies populating the BTFR, low surface brightness (LSB) galaxies are of particular interest, and have been used to investigate the mass distribution and stellar feedback processes at dwarf galaxy scales (e.g. \citealt{zwaan, deblok, dalcantondisks, dicintio2019}). 

Ultra-diffuse galaxies (UDGs; \citealt{vandokkum}) are an especially notable subset of the LSB galaxy population due to their extremely low surface brightness values while having effective radii comparable to L$^{\star}$ galaxies. While these galaxies have been known for decades (e.g. \citealt{sandage}; \citealt{impey}), their recent detection in large numbers in different galaxy clusters, groups, and even in isolated environments (e.g. \citealt{roman2}; \citealt{leisman}; \citealt{paperII}), has sparked a renewed interest in them.

Many UDGs in isolation are H\,{\sc i}--rich, opening the possibility of investigating their gas kinematics. The most systematic study of H\,{\sc i} in UDGs has been carried out by \cite{leisman}, who studied 115 sources\footnote{H\,{\sc i}--rich UDGs represent $\sim$ 6\% of all galaxies with $\mathrm{M_{HI}} \sim$ 10$^{8.8}$ M$_\odot$, with a cosmic abundance similar to cluster UDGs \citep{jones,paperI}.} from the Arecibo Legacy Fast Arecibo L-band Feed Array (\texttt{ALFALFA}) catalogue \citep{alfalfa}, as well as a small subsample of three sources with interferometric H\,{\sc i} data, that meet the optical criteria of having $R_{\rm e} \geq$ 1.5 kpc and $\langle\mu(r,R_{\rm e})\rangle \geq$ 24 mag arcsec$^{-2}$, according to Sloan Digital Sky Survey photometry. 
The authors reported that such galaxies are H\,{\sc i}--rich for their stellar masses and have low star formation efficiencies, similar to other gas-dominated dwarfs (e.g. \citealt{geha}). 
Perhaps most intriguing, \citet{leisman} reported that the velocity widths ($W_{50}$) of the global H\,{\sc i} profiles of their UDGs were significantly narrower than in other \texttt{ALFALFA} galaxies with similar H\,{\sc i} masses. However, without resolved H\,{\sc i} imaging of a significant sample, this result could be attributed to a very strong inclination selection effect for their sample, or systematics when deriving $W_{50}$.


Taking all of the above as a starting point, in this work we undertake 3D--kinematical modeling of resolved H\,{\sc i} synthesis data to study the gas kinematics of six H\,{\sc i}--rich UDGs. The rest of this Letter is organized as follows: in Section \ref{sec:data} we introduce our sample of galaxies with their main properties and we describe our strategy for deriving their kinematics. We present our results and discussion in Section \ref{sec:results}, to then conclude in Section \ref{sec:conclusions}.
Throughout this work we adopt a $\Lambda$CDM cosmology with $\Omega_\textnormal{m}$ = 0.3, $\Omega_{\Lambda}$ = 0.7 and $\mathrm{H_0}$ = 70 km s$^{-1}$ Mpc$^{-1}$.

\section{Sample and kinematics} \label{sec:data}

Our sample consists of six galaxies identified as H\,{\sc i}--bearing UDGs by \cite{leisman}. They have $\mathrm{M_{HI}} \sim$ 10$^9\ $M$_\odot$ and are relatively isolated, by requiring that any neighbor with measured redshift within $\pm$500~km~s$^{-1}$ should be at least at 350 kpc away in projection. Moreover, they have $R_{\rm e} >$ 2 kpc, to ease optical follow-up. 

Our observations were obtained with two interferometers: the data for AGC~122966 and AGC~334315 come from the Westerbork Synthesis Radio Telescope (program R13B/001; PI Adams) and the rest from the Karl G. Jansky Very Large Array (programs 14B-243 and 17A-210; PI Leisman). The observations and data reduction procedure are described in \citet{leisman} and more details will be given in Gault et al. (in prep.). Three more galaxies for which we have data are excluded from this analysis. AGC~238764 seems to have ordered rotation of about 20~km~s$^{-1}$, but our data-cube misses significant flux with respect to the \texttt{ALFALFA} detection. AGC~749251 shows hints of a velocity gradient but it is barely resolved and we are not able to constrain its inclination better than $i \lesssim 30^\circ$. AGC~748738 shows signs of a gradient in velocity but the data are very noisy. We decide not to consider these three galaxies to keep a reliable sample for the kinematic fitting, but more details on these sources will be given in Gault et al. (in prep.).

We estimate the baryonic mass of our UDGs as $\mathrm{M_{bar}}$~=~1.33~$\mathrm{M_{HI}}$~+~$\mathrm{M_\star}$, with $\mathrm{M_{HI}}$ given by:

\begin{equation}
\frac{\mathrm{M_{HI}}}{\mathrm{M_\odot}} = 2.343 \times 10^5 ~ \bigg( \frac{\mathrm{d}}{\textnormal{Mpc}} \bigg)^2 \bigg( \frac{\mathrm{F_{HI}}}{\textnormal{Jy km s$^{-1}$}} \bigg) 
\end{equation}
where we assume (Hubble flow) distances as listed in \citet{leisman}, and fluxes derived from the total H\,{\sc i}--maps using the task \textsc{flux} from \textsc{gipsy} \citep{gipsy2}.

Stellar masses are obtained from the mass-to-light ratio--color relation of \cite{herrmann} for an absolute magnitude in the $g$ band and a $(g-r)$ color. In order to derive such measurements we perform aperture photometry following the procedure described in \citet{antonino} on deep optical data, obtained with the One Degree Imager of the WIYN 3.5-m telescope 
at the Kitt Peak National Observatory (\citealt{leisman}; Gault et al. in prep.). 

\begin{figure*} 
\centering
\includegraphics[scale=0.5263]{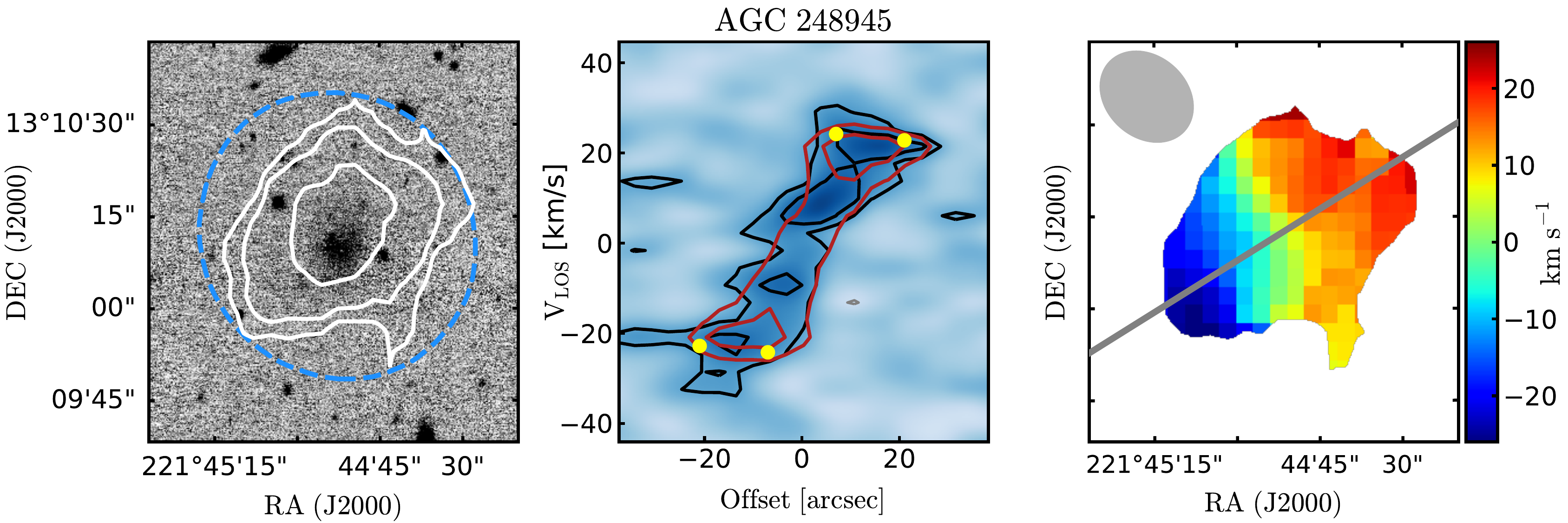}
\caption{A representative galaxy from our sample, AGC~248945. \textit{Left}: H\,{\sc i} contours on top of the $r-$band image; the contours are at 0.88, 1.76 and 3.52 $\times$ 10$^{20}$ H\,{\sc i} atoms per cm$^2$, the outermost contour corresponds to S/N $\approx$ 3. The blue ellipse shows the inclination the galaxy would need to be in the BTFR (see the text for details). \textit{Middle}: PV-diagram along the kinematic major axis; black and red contours correspond to data and $\mathrm{^{3D}}$Barolo best-fit model, respectively; the yellow points show the recovered rotation velocities. \textit{Right}: Observed velocity field, at the same scale as the left panel. The grey line shows the kinematic major axis and the grey ellipse the beam.}
\label{fig:exgalaxy}
\end{figure*}

\begin{figure*} 
    \includegraphics[scale=0.444]{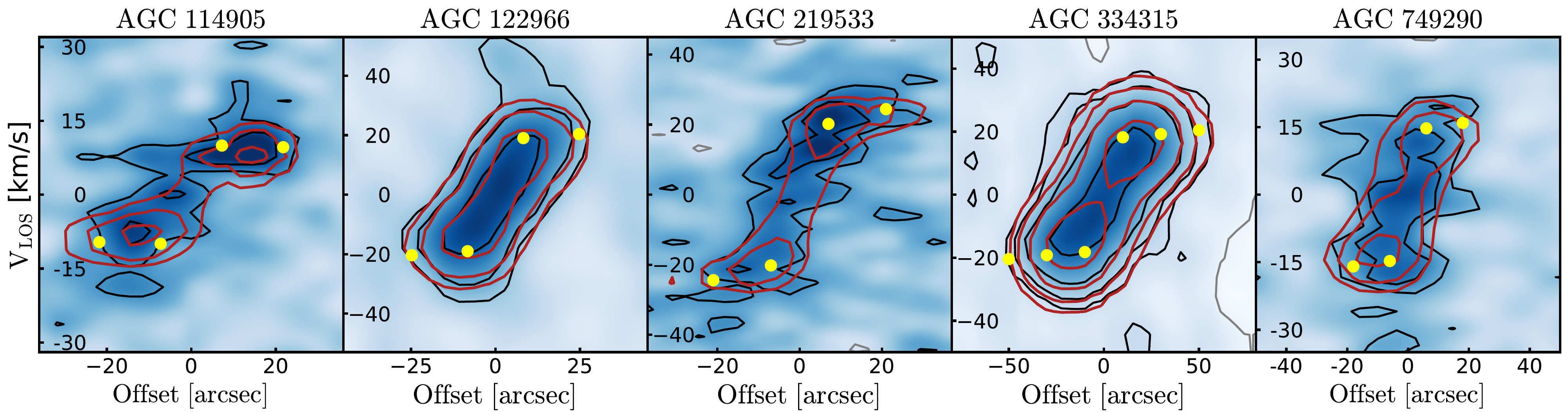}
    \caption{PV slices along the major axes of our galaxies. Contours and points as in Figure~\ref{fig:exgalaxy}, where AGC~248945 is shown. The narrowness of the PV diagrams suggests low gas velocity dispersions, as confirmed by $\mathrm{^{3D}}$Barolo.}
    \label{fig:pvs}
\end{figure*}

We find a mean $\mathrm{M_{HI}\ /\ M_\star} \approx$ 15, confirming that the baryonic budget is mainly set by the H\,{\sc i} content, which we can robustly measure. Table~\ref{tab:sample} gives the name, distance, inclination, baryonic mass, gas-to-stellar mass ratio, circular velocity, central surface brightness and color of our galaxies. Figure \ref{fig:exgalaxy} shows the stellar image, 0$^{\rm th}$-moment map, major-axis position-velocity (PV) diagram, and observed velocity field for a representative case, AGC~248945. Figure~\ref{fig:pvs} shows the PV diagrams for the rest of our sample.

\begin{deluxetable*}{lccccccc}
\tablecaption{Name, distance, inclination, baryonic mass, gas-to-stellar mass ratio, circular velocity, central surface brightness and color of our sample.}
\tablehead{ \noalign{\smallskip}
\colhead{Name} & \colhead{Distance} & \colhead{Inclination} & \colhead{$\mathrm{log(M_{bar}/M_\odot)}$} & \colhead{$\mathrm{M_{gas}/M_{\star}}$} & \colhead{$\mathrm{V_{circ}}$} & \colhead{$\mathrm{\mu(g,0)}$} & \colhead{$g-r$} \\
\colhead{} & \colhead{(Mpc)} & \colhead{(deg)} & \colhead{} & \colhead{} & \colhead{(km s$^{-1}$)} & \colhead{(mag~arcsec$^{-2}$)} & \colhead{(mag)}
}
\startdata
     AGC 114905 &  76 & 33 & 9.21 $\pm$ 0.20 & 7.1$^{+4.9}_{-2.3}$ & 19$^{+6}_{-4}$ & 23.62 $\pm$  0.13 &  0.30 $\pm$ 0.12    \\ \noalign{\smallskip}  
     AGC 122966 &  90 & 34 & 9.21 $\pm$ 0.14 & 29.1$^{+11.9}_{-7.0}$ & 37$^{+6}_{-5}$ & 25.38 $\pm$  0.23 & -0.10 $\pm$ 0.22    \\ \noalign{\smallskip}
     AGC 219533 &  96 & 42 & 9.36 $\pm$ 0.27 & 19.7$^{+12.2}_{-8.8}$ & 37$^{+5}_{-6}$ & 24.07 $\pm$  0.33 &  0.12 $\pm$ 0.12    \\ \noalign{\smallskip}
     AGC 248945 &  84 & 66 & 9.05 $\pm$ 0.20 & 2.4$^{+1.6}_{-0.8}$ & 27$^{+3}_{-3}$ & 23.32 $\pm$  0.35 &  0.32 $\pm$ 0.11    \\ \noalign{\smallskip}
     AGC 334315 &  73 & 52 & 9.32 $\pm$ 0.14 & 23.7$^{+9.8}_{-5.9}$ & 26$^{+4}_{-3}$ & 24.52 $\pm$  0.13 & -0.08 $\pm$ 0.18    \\ \noalign{\smallskip}
     AGC 749290 &  97 & 39 & 9.17 $\pm$ 0.17 & 6.1$^{+2.9}_{-1.7}$ & 26$^{+6}_{-6}$ & 24.66 $\pm$  0.30 &  0.17 $\pm$ 0.12    \\ \noalign{\smallskip}
\enddata \label{tab:sample}
\tablecomments{Distances, taken from \citet{leisman}, have an uncertainty of $\pm$5 Mpc, while the uncertainty for the inclination is ${\pm 5^\circ}$. The central surface brightness is obtained from an exponential fit to the $g-$band surface brightness profile.}
\end{deluxetable*}

{}


Rotation velocities are derived with the software $\mathrm{^{3D}}$Barolo\footnote{Version 1.4, \url{{ http://editeodoro.github.io/Bbarolo/}}} \citep{barolo}, which fits tilted-ring disc models to the H\,{\sc i} data-cubes (e.g. \citealt{iorio}; \citealt{ceci}). This approach is particularly suited to deal with our low spatial resolution data ($2-3$ resolution elements per galaxy side) as it is virtually unaffected by beam-smearing (e.g. \citealt{enrico2}).
While further details about the properties of our sample and the configuration used in $\mathrm{^{3D}}$Barolo will be given in Mancera Pi\~na et al. (in prep.), here we briefly summarize our methodology. 

We give the position angle and inclination to $\mathrm{^{3D}}$Barolo. For the former we choose the angle that maximizes the amplitude of the PV slice along the major axis. 
The inclination of each galaxy is derived by minimizing the residuals between its observed 0$^{\rm th}$-moment map and the 0$^{\rm th}$-moment map of models of the same galaxy projected at different inclinations between $10^\circ - 80^\circ$.
We have tested this method blindly, without a priori knowledge of the position angle, inclination nor rotation velocity, on a sample of 32 H\,{\sc i}--rich dwarfs drawn from the \textsc{apostle} cosmological hydrodynamical simulations \citep{apostle2,apostle}, from which mock data-cubes have been produced at resolution and S/N matching our observations, using the {\sc martini} software\footnote{Version 1.0.2, \url{http://github.com/kyleaoman/martini}} \citep{kyle2019}. We find that we can consistently recover the position angle within $\pm 8^\circ$ and the inclination within $\pm 5^\circ$ as long as $i\gtrsim 30^\circ$, with no systematic trends. These small uncertainties in position angle and inclination have no significant impact on the recovered rotation velocities.

We run $\mathrm{^{3D}}$Barolo with fixed inclination and position angle, and the rotation velocity and velocity dispersion as free parameters, for our fiducial inclination $i$, as well as for $i+5^{\circ}$ and $i-5^{\circ}$. We find rotation velocities ($\mathrm{V_{rot}}$) suggesting flat rotation curves for all our sample. For calculating $\mathrm{V_{rot}}$, we use the mean velocity of the rings, as found with our fiducial inclination. The associated uncertainties come from the 16$^{\rm th}$ and 84$^{\rm th}$ percentiles of the velocity distribution obtained when considering the uncertainty in our inclination.
To convert from $\mathrm{V_{rot}}$ to circular velocity ($\mathrm{V_{circ}}$), we correct for pressure supported motions using $\mathrm{^{3D}}$Barolo as well (cf. \citealt{iorio}). As suggested by the narrowness of the PV diagrams (Fig.~\ref{fig:exgalaxy} and \ref{fig:pvs}), we find low velocity dispersions (Mancera Pi\~na et al. in prep.), giving rise to very small asymmetric drift corrections ($\lesssim$~2~km~s$^{-1}$).


\section{Results and discussion} \label{sec:results}

\begin{figure*}[t!]
    \centering
    \includegraphics[scale=0.62]{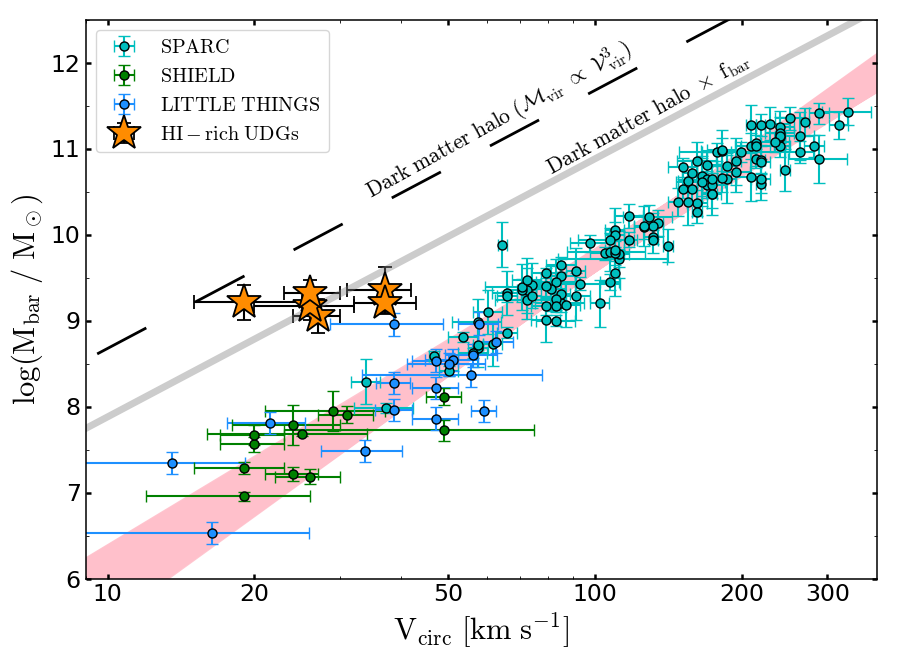}
    \caption{Circular velocity versus baryonic mass plane. Galaxies from the SPARC, SHIELD and LITTLE THINGS samples lie on top of the BTFR. The pink area is the 99\% confidence interval of an orthogonal distance regression to the SPARC sample.} H\,{\sc i}--rich UDGs are clear outliers of the BTFR, and in a position consistent with having no ``missing baryons".
    \label{fig:BTF}
\end{figure*}

In Figure \ref{fig:BTF} we present the circular velocity--baryonic mass plane for our H\,{\sc i}--rich UDGs, compared with galaxies from the SPARC \citep{sparc}, SHIELD \citep{shield} and LITTLE THINGS \citep{iorio} samples. Clearly, all the UDGs studied here lie significantly above the BTFR.

Our galaxies rotate about 3 times lower than galaxies with comparable $\mathrm{M_{bar}}$ and effective radius (but higher surface brightness). Alternatively, they have about 10--100 times the $\mathrm{M_{bar}}$ of galaxies with similar $\mathrm{V_{circ}}$ (but smaller effective radius and higher surface brightness, on average). These low velocities are consistent with the observations by \citet{leisman} and \citet{jano} of H\,{\sc i}--rich UDGs having narrower $W_{50}$ than galaxies of similar H\,{\sc i} mass.

Before discussing the implications of this result we address its robustness. The baryonic masses here derived cannot be substantially overestimated: H\,{\sc i} line fluxes can be measured with good accuracy (and we find fluxes in agreement with those derived from \texttt{ALFALFA} data by \citealt{leisman}), and the distances to the galaxies in our sample ($\langle \mathrm{d} \rangle \sim$ 90 Mpc) are large enough to be well represented by Hubble flow models, so the estimation of their H\,{\sc i} mass is reliable. The H\,{\sc i}--rich nature of our galaxies also implies that the stellar mass and its systematics play a rather minor role: even M$_\star$~=~0 would not move the galaxies significantly in Figure~\ref{fig:BTF}.

A severe underestimation of the rotation velocities is also unlikely. First, the H\,{\sc i} emission of the galaxies extends out to radii $\approx$ 8--18 kpc, and velocities obtained at such large radii are expected to be tracing the maximum of the rotation curve for any plausible dwarf galaxy dark matter halo (e.g. \citealt{kyle2015}, their Fig.~2).
Second, regarding the inclination correction, bringing the galaxies back to the BTFR would require a nearly face-on inclination ($i \approx 10^{\circ}-20^{\circ}$) for $all$ of them, which is both unlikely and incompatible with the observed intensity maps, as illustrated in Figure~\ref{fig:exgalaxy}, with an ellipse showing the inclination that the galaxy would need to be on the BTFR.
Third, non-circular motions are not strong enough to solve the observed discrepancy: regardless of the mode(s), their order, phase or amplitude, harmonic non-circular motions do not bias $\mathrm{V_{rot}}$ towards lower values systematically, as long as the viewing angle of the galaxy is random (\citealt{kyle2019}, their Fig.~7), and the symmetry of the approaching and receding sides of our PV-diagrams suggests the absence of anharmonic components. We also investigated with $\mathrm{^{3D}}$Barolo the presence of radial motions, but no clear evidence for this was found, although higher-resolution observations are needed to further confirm this.

Finally, it is worth to mention that the observed velocity gradients cannot be attributed to H\,{\sc i} winds: in that case the gas velocity dispersion would be much higher than observed, and the galaxies would need very high star formation rate densities, opposite to what is measured \citep{leisman}.\\


\noindent
Previous studies already suggested the existence of outliers in the BTFR, or at least an increase in its scatter at low $\mathrm{V_{circ}}$ (e.g. \citealt{geha}). Sometimes, however, the robustness of the measurements of the rotation velocities (usually estimated from the global H\,{\sc i} profile) and inclinations of such outliers were unclear (cf. \citealt{kyle2016} and references therein).

\noindent
Based on the discussion above, we conclude that the positions of H\,{\sc i}--rich UDGs in the $\mathrm{M_{bar}-V_{circ}}$ plane derived here are robust, and our UDGs do not follow the BTFR\footnote{It is worth to notice that the two outliers close to our UDGs, DDO 50 and UGC 7125, also have relatively large effective radii and/or low surface brightness.}. 
This suggests that the distribution of late-type systems in such plane is broader than previously observed, and may have important implications for the scatter in the BTFR, which is a strong constraint for cosmological models. Despite the small scatter previously reported (e.g. \citealt{lelli2016,anastasia}), our findings open the possibility for a scenario where the parameter space in the $\mathrm{M_{bar}-V_{circ}}$ plane between the UDGs presented here and the BTFR is populated by LSB galaxies whose resolved H\,{\sc i} kinematics have not been studied yet, and which are not in our sample due to sharp selection effects. This may increase the error budget of the intrinsic scatter of the relation, but to properly understand the magnitude of this effect a more complete census of the relative abundances of these galaxies is required.\\

  


\noindent
A second result emerges when comparing the position of our galaxies with the curves in Figure~\ref{fig:BTF}. The black dashed curve is the relation between the circular velocity at the virial radius and the virial mass of dark matter haloes ($\mathrm{M_{vir}/M_\odot}~\approx~4.75~\times~10^{5}~\mathrm{(V_{vir}/km~s^{-1})^3}$, for $\mathrm{\Delta_c = 100}$, cf. \citealt{mcgaugh3}). If $\mathrm{M_{vir}}$ is multiplied by the cosmological baryon fraction ($\mathrm{f_{bar}} \approx$ 0.16), this gives rise to the solid grey curve, indicating the expected position for galaxies with a baryon fraction equal to $\mathrm{f_{bar}}$\footnote{Note that this assumes $\mathrm{V_{circ} \approx V_{vir}}$, but in general $\mathrm{V_{circ}}$ tends to be slightly larger for massive galaxies ($\mathrm{V_{circ} \approx 1.5V_{vir}}$). This would flatten the grey curve at high $\mathrm{V_{circ}}$ values.}. 
Unexpectedly, our UDGs lie on top this curve, meaning that they are consistent with having no ``missing baryons". 


\citet{postinonmissing} recently discovered that some massive spirals have virtually no ``missing baryons". There is, however, a substantial difference between our UDGs and these massive spirals, as the former are H\,{\sc i}--dominated and have very shallow potential wells compared to the latter. How, then, is it possible that they retained all of their gas? One intriguing possibility is that they have not experienced strong episodes of gas ejection: feedback processes must have been relatively weak and the shallow gravitational potentials managed to retain (or promptly re-accrete) all of their baryons. We surmise that this could be related to the low gas velocity dispersions we find for our sample, which suggest a currently weak heating of the gas. This may be analogous to the ``failed feedback problem" of \citet{postinonmissing}, although in their case feedback has failed at limiting the star formation efficiency of massive spiral galaxies.

Extremely efficient feedback has been invoked to solve different discrepancies between observations and $\Lambda$CDM predictions (see \citealt{tulin} and \citealt{bullock} for a review, including limitations of such solutions), as well as to explain the formation of UDGs via feedback-driven outflows resulting from bursty star formation histories (e.g. \citealt{dicintio2017}). These new observations seem to present a challenge to these models. 

An alternative scenario could be that our galaxies reside in haloes with $\mathrm{V_{circ}~\approx~80~km~s^{-1}}$ but very low concentration, such that their rotation curves are still rising at our outermost measured radii. However, this does not seem feasible since the concentration parameter needed for this is $c~\approx$ 1, instead of the expected $c~\approx 10$ \citep{concentration}, making the existence of such galaxies within the volume of the Universe basically impossible.\\

\begin{figure} 
    \centering
    \includegraphics[scale=0.455]{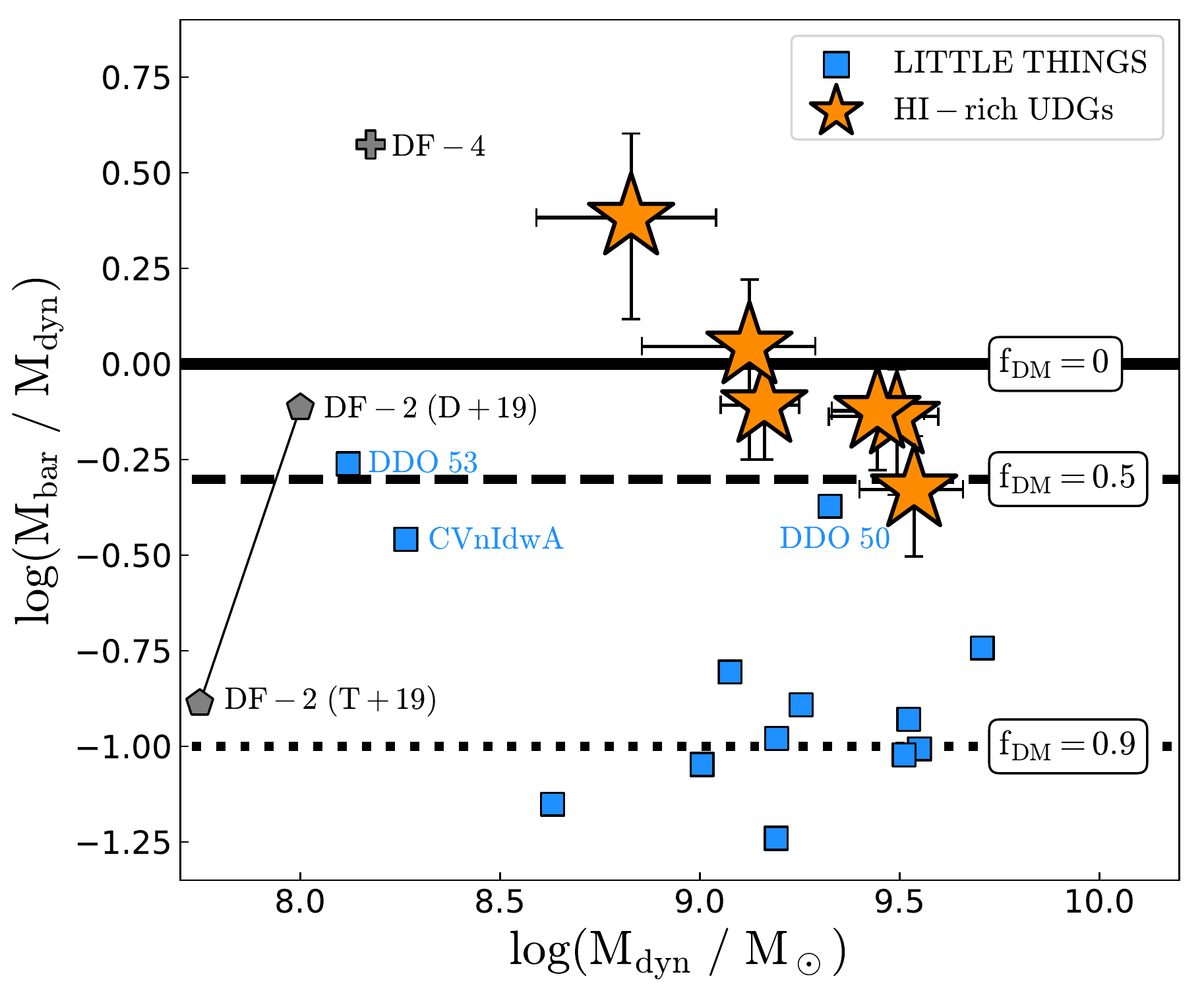}
    \caption{Baryonic to dynamical mass ratio as a function of the dynamical mass, measured inside $\approx~4~\mathrm{R_{\rm d}}$. The solid, dashed and dotted lines show the position where galaxies with 0\%, 50\% and 90\% dark matter lie, respectively. LITTLE THINGS galaxies \citep{iorio} are shown for comparison, as well as two estimates for DF--2 (\citealt{danieli}, D+19 and \citealt{trujilloDF2}, T+19) and DF--4 \citet{vandokkumDM2}, for which we assume $\mathrm{M_{bar}} = \mathrm{M_\star}$.}
    \label{fig:DM}
\end{figure}

\noindent
Figure \ref{fig:DM} shows the ratio between baryonic and dynamical mass of our UDGs, with a dynamical mass estimated as $\mathrm{M_{dyn}(<R_{out}) = V_{circ}^2~R_{out} / G}$, with $\mathrm{R_{out}}$ the radius of the outermost point of the rotation curve. Both our sample and LITTLE THINGS galaxies have a mean $\mathrm{R_{out}/R_{\rm d}} \approx 4$, with $\mathrm{R_{\rm d}}$ the optical disc-scale length. 

Even if our H\,{\sc i}--rich UDGs have a baryon fraction equal to the cosmological average, their dynamics could be dark matter-dominated at all radii, as other galaxies of similar $\mathrm{V_{circ}}$, but this is does not seem to be the case, since $\mathrm{M_{bar}(R<R_{out}) \approx M_{dyn}(R<R_{out})}$. Although more precise values of $\mathrm{M_{bar}}$ and $\mathrm{M_{dyn}}$ should be determined with better data, Figure~\ref{fig:DM} indicates that these galaxies have much less dark matter within the extent of their discs than other dwarfs and LSB galaxies, and that, inside their discs, the baryonic component dominates.

The dynamical properties here shown resemble those of tidal dwarf galaxies \citep{hunterTD, lelliTD}. However, given the isolation (mean distance to nearest neighbor $\sim$ 1 Mpc) of our UDGs, a tidal dwarf origin does not seem likely, but this is hard to test with the current data.

Based on their globular clusters kinematics the UDGs NGC1052-DF2 \citep{vandokkumDM, danieli} and NGC1052-DF4 \citep{vandokkumDM2} have recently been claimed to lack dark matter, although some concerns exist regarding their distances and environments \citep{trujilloDF2, monelli}. Our UDGs have robust distances determined from their recession velocities and avoid dense environments, mitigating these concerns. They may be subject to different systematics, but demonstrate that there may indeed exist a previously under-appreciated population of unusually dark matter-deficient galaxies.



\section{Conclusions} \label{sec:conclusions}
We have analyzed a set of interferometric H\,{\sc i} line observations of gas--dominated UDGs. Using a 3D fitting technique we obtain robust measurements of their circular velocities, allowing us to place them in the circular velocity--baryonic mass plane. 

We find that our six galaxies lie well above the BTFR, with rotation velocities too low given their baryonic masses. 
Their position in the circular velocity--baryonic mass plane implies that they have a baryon fraction within their virial radius equal or close to the cosmological value, and we speculate that this could be due to extremely inefficient feedback, challenging our current understanding of feedback processes in dwarfs. Additionally, the dynamics of these galaxies are dominated by the baryonic component out to the outermost measured radii, and they have very low dark matter fractions inside such radii, suggesting a broader distribution in the dark matter content of galaxies than previously thought.

The fact that galaxies with these properties had not been reported before is perhaps because interferometric H\,{\sc i} observations are usually targeted based on previous optical studies. Since UDGs are an extremely optically faint population, it is not particularly surprising that this galaxy population has not been identified before. With the advent of large H\,{\sc i} interferometric surveys we expect this hidden population to come to light.


\vspace*{5ex}
\acknowledgments
\small
We appreciate the careful revision and useful comments made by an anonymous referee.
We thank Giuliano Iorio and Andrew McNichols for their clarifications on LITTLE THINGS and SHIELD data, respectively. We would also like to thank Anastasia Ponomareva, Arianna Di Cintio and Federico Lelli for interesting discussions. 

PEMP and FF are supported by the Netherlands Research School for Astronomy (Nederlandse Onderzoekschool voor Astronomie, NOVA), Phase-5 research programme Network~1, Project~10.1.5.6. EAKA is supported by the WISE research programme, which is financed by the Netherlands Organization for Scientific Research (NWO). KAO received support from VICI grant 016.130.338 of NWO. LP acknowledges support from the Centre National d'\'{E}tudes Spatiales (CNES). MPH is supported by grants NSF/AST-1714828 and from the Brinson Foundation. This work has been supported in part by NSF grant AST-1625483 to KLR, and by The National Radio Astronomy Observatory (The National Radio Astronomy Observatory is a facility of the National Science Foundation operated under cooperative agreement by Associated Universities, Inc.). We have made an extensive use of SIMBAD and ADS services, for which we are thankful.







\begin{thebibliography}{999}

  
  \bibitem[Bacchini et al.(2018)]{ceci} Bacchini, C., Fraternali, F., Iorio, G., Pezzulli, G., 2019, A\&A, 622, 64
  
  
  
  
  \bibitem[Bullock \& Boylan-Kolchin(2017)]{bullock} Bullock J. S., Boylan-Kolchin M., 2017, ARAA, 55, 343
  
  


 
      \bibitem[Dalcanton et al.(1997)]{dalcantondisks} Dalcanton, J. et al. 1997, ApJ, 482, 659
      
    \bibitem[den Heijer et al.(2015)]{heijer} den Heijer M. et al. 2015, A\&A, 581, A98
      
      
      \bibitem[Danieli et al.(2019)]{danieli} Danieli S., van Dokkum P., Conroy C., Abraham R., RomanowskyA. J., 2019, ApJ, 874, 12
      
       \bibitem[de Blok(1997)]{deblok} de Blok, W. J. G. 1997, PhD thesis, University of Groningen
      
    \bibitem[Di Cintio et al.(2017)]{dicintio2017} Di Cintio, A. et al. 2017, MNRAS, 466, 1

    \bibitem[Di Cintio et al.(2019)]{dicintio2019} Di Cintio, A. et al. 2019, MNRAS, 486, 2535
    
          \bibitem[Di Teodoro \& Fraternali(2015)]{barolo} Di Teodoro, E. \& Fraternali, F., 2015, MNRAS, 451, 3021
      
  \bibitem[Di Teodoro et al.(2016)]{enrico2} Di Teodoro, E. M., Fraternali, F., \& Miller, S. H. 2016, A\&A, 594, A77

\bibitem[Dutton(2012)]{dutton} Dutton, A. A., 2012, MNRAS, 424, 3123


\bibitem[Fattahi et al.(2016)]{apostle2} Fattahi, A., Navarro, J. F., Sawala, T. et al. 2016, MNRAS,457, 844

\bibitem[Geha et al.(2006)]{geha} Geha, M., Blanton, M. R., Masjedi, M., \& West, A. A. 2006, ApJ, 653, 240

\bibitem[Giovanelli et al.(2005)]{alfalfa} Giovanelli, R., Haynes, M. P., Kent, B. R. et al. 2005, AJ, 130, 2598

\bibitem[Governato et al.(2007)]{governato} Governato, F. et al. 2007, MNRAS, 374, 1479


\bibitem[Herrmann et al.(2016)]{herrmann} Herrmann K. A., Hunter D. A., Zhang H.-X., Elmegreen B. G., 2016, AJ, 152, 17

\bibitem[Hunter et al.(2000)]{hunterTD} Hunter, D. A., Hunsberger, S. D., \& Roye, E. W. 2000, ApJ, 542, 137

  
 
  \bibitem[Iorio et al.(2017)]{iorio} Iorio G., Fraternali F., Nipoti C., Di Teodoro E., Read J. I., Battaglia G., 2017, MNRAS, 466, 415
  
 \bibitem[Impey et al.(1988)]{impey} Impey C., Bothun G., Malin D. 1988, ApJ, 330, 634

  
  \bibitem[Janowiecki et al.(2019)]{jano} Janowiecki, S. et al. 2019, MNRAS, DOI: 10.1093/mnras/stz1868
  
      \bibitem[Jones et al.(2018)]{jones} Jones, M. G. et al. 2018, A\&A, 614, 21

    \bibitem[Leisman et al.(2017)]{leisman} Leisman L. et al. 2017, ApJ, 842, 13

\bibitem[Lelli et al.(2015)]{lelliTD} Lelli, F. et al. 2015, A\&A, 584, A113

\bibitem[Lelli et al.(2016a)]{lelli2016} Lelli, F., McGaugh, S. S., Schombert, J. M., 2016a, ApJ, 816, 14

 \bibitem[Lelli et al.(2016b)]{sparc}  Lelli F., McGaugh S. S., Schombert J. M., 2016b, AJ, 152, 15


\bibitem[Ludlow et al.(2014)]{concentration} Ludlow A. D. et al. 2014, MNRAS, 441,378

\bibitem[Mancera Pi\~na et al.(2019)]{paperII} Mancera Pi\~na, P. E., Aguerri, J.A.L., Peletier, R.F., Venhola, A., Trager, S., Choque Challapa, N., 2019, MNRAS, 485, 1036

   \bibitem[Mancera Pi\~na et al.(2018)]{paperI} Mancera Pi\~na, P. E., Peletier, R. F., Aguerri, J.A.L., Venhola, A., Trager, S., Choque Challapa, N., 2018, MNRAS, 481, 4381

\bibitem[Marasco et al.(2019)]{antonino} Marasco, A., Fraternali, F., Posti, L., Ijtsma, M., Di Teodoro, E. M., Oosterloo, T. 2019, A\&A, 621, 6



\bibitem[McGaugh et al.(2005)]{mcgaugh2} McGaugh, S. S., 2005, ApJ, 632, 859

\bibitem[McGaugh(2012)]{mcgaugh3} McGaugh, S. S., 2012, AJ, 143, 40

\bibitem[McGaugh et al.(2000)]{mcgaugh1} McGaugh, S. S., Schombert, J. M.; Bothun, G. D.; de Blok, W. J. G., 2000, ApJ, 533, 99

 \bibitem[McNichols et al.(2016)]{shield} McNichols, A. T. et al. 2016, ApJ, 832, 89
 

\bibitem[Monelli \& Trujillo(2019)]{monelli} Monelli, M. \& Trujillo, I., 2019, ApJ, 880, 11



  
  \bibitem[Oman et al.(2015)]{kyle2015} Oman, K. et al. 2015, MNRAS, 452, 3650
  
  \bibitem[Oman et al.(2016)]{kyle2016} Oman, K., Navarro, J. F., Sales, L. V. et al. 2016, MNRAS, 460, 3610
  
  \bibitem[Oman et al.(2019)]{kyle2019} Oman, K. et al. 2019, MNRAS, 482, 8210
  
     
 
     \bibitem[Ponomareva et al.(2017)]{anastasia} Ponomareva  A.  A.,  Verheijen  M.  A.  W.,  Peletier  R.  F.,  Bosma A., 2017, MNRAS, 469, 2387
     
 \bibitem[Posti et al.(2019)]{postinonmissing} Posti, L., Fraternali, F. \& Marasco, A., 2019, A\&A, 626, 56
 



    \bibitem[Rom\'an \& Trujillo(2017)]{roman2} Rom\'an, J., \& Trujillo, I. 2017, MNRAS, 468, 4039
    
    \bibitem[Sales et al.(2017)]{sales} Sales L. V. et al. 2017, MNRAS, 464, 2419
    
    \bibitem[Sandage \& Binggeli(1984)]{sandage} Sandage \& Binggeli, 1984, AJ, 89, 919
    
    \bibitem[Sawala et al.(2016)]{apostle} Sawala, T., Frenk, C. S., Fattahi, A. et al. 2016, MNRAS, 457, 1931
    
    
    
  
    
    
    \bibitem[Trujillo et al.(2019)]{trujilloDF2} Trujillo, I. et al. 2019, MNRAS, 486, 1192
    
    \bibitem[Tulin \& Yu(2018)]{tulin} Tulin S., Yu H.-B., 2018, Phys. Rep., 730


   \bibitem[van Dokkum et al.(2015)]{vandokkum} van Dokkum, P. G., Abraham, R., Merritt, A. et al. 2015, ApJL, 798,  L45

\bibitem[van Dokkum et al.(2018)]{vandokkumDM} van Dokkum, P. et al. 2018, Nature, 555, 629

\bibitem[van Dokkum et al.(2019)]{vandokkumDM2} van Dokkum, P. et al. 2019, ApJL, 874, 5




\bibitem[Vogelaar \& Terlouw(2001)]{gipsy2} Vogelaar M. G. R., Terlouw J. P., 2001, in Harnden Jr. F. R., Primini F. A., Payne H. E., eds, Astronomical Society of the Pacific Conference Series Vol. 238, Astronomical Data Analysis Software and Systems X. p. 358



\bibitem[Zwaan et al.(1995)]{zwaan} Zwaan, M., van der Hulst, J., de Blok, W. \& McGaugh, S., 1995, MNRAS, 273, L35
\end{thebibliography}
\end{document}